\documentclass[twocolumns]{aa} 
\usepackage{graphicx}
\usepackage{amsmath}
\usepackage{epstopdf}
\usepackage{grffile}
\usepackage{amssymb}
\usepackage{lscape}
\usepackage{txfonts}
 \usepackage{natbib}
\begin{document}

\newcommand{\dd}{deg$^{2}$}
\newcommand{\flux}{$\rm erg \, s^{-1} \, cm^{-2}$}
   \title{A Chandra view of  the z=1.62 galaxy cluster IRC-0218A }


   \author{M. Pierre
          \inst{1}\thanks{mpierre@cea.fr}
          \and
          N. Clerc \inst{1}
          \and
          B. Maughan \inst{2}
          \and
          F. Pacaud \inst{3}
          \and
          C. Papovich \inst{4}
          \and
          C. N. A. Willmer \inst{5}
          }

   \institute{Laboratoire AIM, CEA/DSM/IRFU/SAp, CEA Saclay, 91191 Gif-sur-Yvette, France 
         \and
             HH Wills Physics Laboratory, University of Bristol, UK 
          \and
                Argelander-Institut f\"ur Astronomie, University of Bonn, Auf dem H\"ugel 71, 53121 Bonn, Germany  
          \and
Texas A\&M, University 
Department of Physics and Astronomy,
College Station, TX 77843-4242
\and
Steward Observatory, University of Arizona, 933 N.
Cherry Avenue, Tucson, AZ, 85721, USA
                        }

   \date{Received September 28th, 2011;  December 13th, 2011}

 
  \abstract
   {Very few $z>1.5$ clusters of galaxies are currently known. It is important to study the properties of  galaxies in these clusters and the intra-cluster medium and, furthermore, to cross-check the reliability of the various mass estimates. This will help  to clarify the process of structure formation and how distant clusters may be used to constrain cosmology.}
   {We present a 84 ks Chandra observation of IRC-0218A, a cluster of galaxies inferred by the presence of a galaxy overdensity in the infrared at a redshift of 1.62 and associated with some XMM emission.}
   {We performed a spatial analysis of the Chandra X-ray photon distribution. }
   {The Chandra observation of IRC-0218A  appears to be entirely dominated by a point source located at the centroid of the mid-infrared galaxy density. In addition, we detected  weak extended emission ($2.3\sigma$)  out to a radius of $25''$ with a flux of    $\sim 3 ~10^{-15}$ \flux\ in the [0.3-2] keV band. Assuming that clusters evolve similarly,  we infer a virial mass of  $M_{200}= 7.7(\pm 3.8)~10^{13}M_{\odot}$. This is marginally compatible with our current estimate of the cluster dynamical mass (based on 10 redshifts), although there is no evidence that the galaxy peculiar velocities correspond to the motions of a virialized structure. The stellar mass enclosed in the inferred X-ray virial radius is estimated to be 1-2$~10^{12}M_{\odot}$. \\         We provide a detailed account of 28 X-ray point sources detected in the field.   }
   {}


   \maketitle
%

\section{Introduction}

Distant massive clusters of galaxies are, in theory, key objects for constraining cosmology because their abundance strongly depends  on $\sigma_{8}$ and $ \Omega_{m}$. To second order, they are quite sensitive to the equation of state of the dark energy and to possible non-Gaussian features in the initial spectrum of density perturbations. However, the quest for and the study of distant massive clusters is a tedious task, since these objects are expected to be rare (no more than $\sim$ two Coma-type clusters are expected beyond $z>1$ over the whole sky) and, to date, only a tiny fraction of the distant universe has been investigated at a sufficient sensitivity. Another difficulty in involving distant clusters in cosmological studies is that it is a priori very difficult to properly estimate their mass: X-ray cluster scaling laws are  now rather well assessed in the local universe, but their evolution is still a matter of debate, mainly because of the difficulty in assembling unbiased samples of distant clusters; one may derive a dynamical or X-ray mass estimates under the assumption  that the system is well virialized - which becomes a challenging hypothesis at high redshift; weak lensing mass determinations are increasingly  hampered by projection effects with increasing redshift - moreover, they lack the necessary sensitivity beyond $z>1.5$.

While, by definition, a cluster of galaxies is a collection of galaxies bound in a common potential well, hence having similar recession velocities, clusters are usually unambiguously identified by the presence of extended emission  from the hot gas trapped in the cluster potential; this is essentially because projection effects can always mimic  the presence of a cluster to some extent, even in the velocity space, while significant X-ray emission is only possible from a gas that is denser than the one expected to reside in the cosmic filaments ($>  10^{-4}/cm^{3}$).

At present, only four clusters with spectroscopic redshifts are known beyond $z>1.5$. Two of them have been primarily detected in the X-ray waveband \citep{fassbender11, santos11}. The other two were identified through an overdensity of red galaxies; weak X-ray emission was found a posteriori to be associated with both of them \citep{papovich10, gobat11}.

Cluster IRC-0218A was almost simultaneously identified by \cite{papovich10} using Spitzer data and secured by  optical spectroscopic observation and by \cite{tanaka10} using deep multi-band photometry along with near-IR spectroscopy.  \cite{papovich10} provided redshifts for five (seven)  blue galaxies with $1.62 < z < 1.63$ and r$ < 1 (1.5)$ Mpc.   \cite{tanaka10}  provided redshifts for K-band selected objects and had some red galaxies (the authors did not provide a redshift table, but only a finding chart). The presence of a possible companion cluster some two arcmin east of the main clump was  also pointed out by \cite{tanaka10}.  Cluster IRC-0218A  is located in the deepest part of the XMM-LSS survey, the Subaru Deep Survey, and hence X-ray information was readily available for this cluster. Indeed, both authors report the presence of some X-ray emission associated with the cluster \citep[and with the companion, for][] {tanaka10}. The object, however, happens to lie at the very edge of three adjacent XMM observations, at an off-axis of $\sim 12'$, preventing a proper characterization of the X-ray emission.  
We have thus obtained a deep Chandra observation to examine the properties of the X-ray emission associated to IRC-0218A in detail.
 
In Section 2 we describe the Chandra observation.  Sec. 3 presents the spatial analysis of the Chandra emission. Implications for the cluster mass estimate are discussed in Sec. 4.  
Throughout the article, we assume the WMAP5 cosmology \citep{dunkley09} for which $1'' = 8.6$ kpc.

\section{The Chandra observation}
\begin{figure}
   \centering
    \includegraphics[width=8cm]{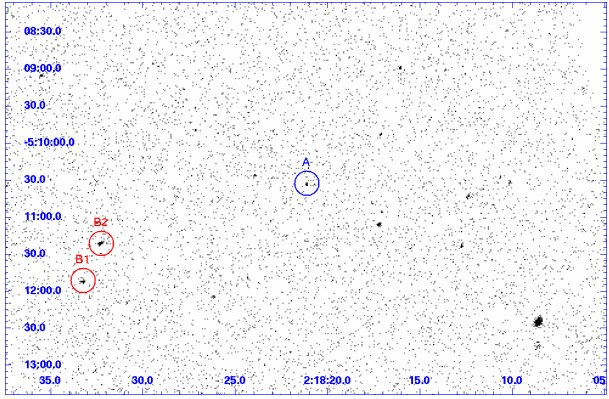}  
   \caption{Unbinned Chandra photon-image in the [0.5-2] keV band of cluster IRC-0218A  for a total exposure time of $\sim$ 84 ks. The blue circle indicates the position of the main cluster component (A) as inferred from the  Spitzer and XMM data. The two red circles (B1, B2) indicate point sources in the vicinity of a possible second cluster component  as proposed by \cite{tanaka10}. }
              \label{cluster-image}
    \end{figure}

\begin{figure*}
   \centering      
        \includegraphics[width=18.5cm]{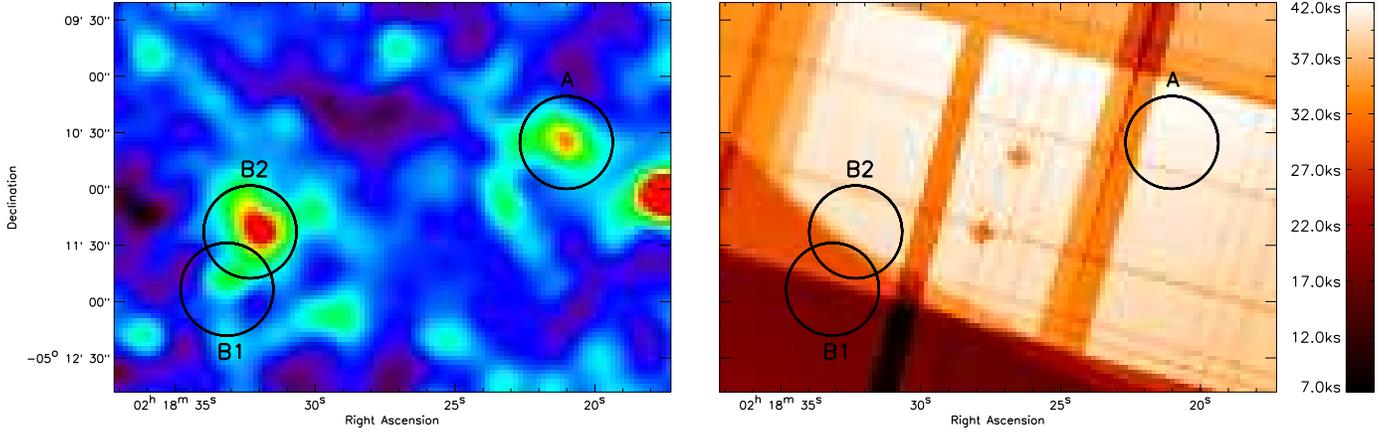}
   \caption{The XMM image of IRC-0218A is a composite of three adjacent observations with a resulting mean off-axis distance of $\sim 12' $. Left (a): the photon image adaptively smoothed (boxcar smoothing with minimum threshold of 10 photons).  The position of the three Chandra point sources (Fig. \ref{cluster-image}) is indicated by circles having a radius of $25''$.  Right (b): The merged exposure map; the color scale indicates the cumulative effective exposure.}
              \label{xmm-image}
    \end{figure*}

\begin{figure*}
   \centering      
        \includegraphics[width=18cm]{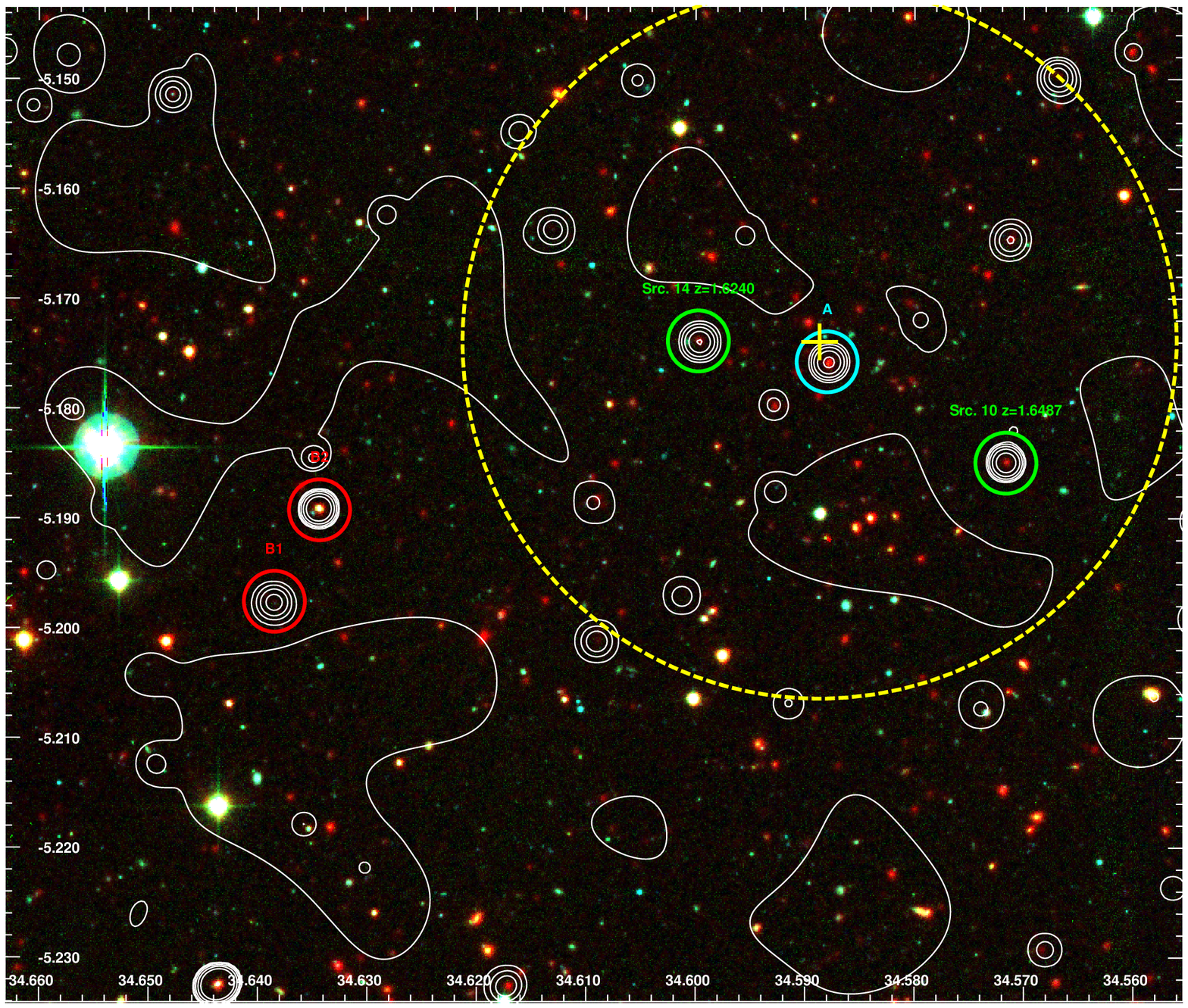}
   \caption{Chandra contours in the [0.5-2] keV band, overlaid on a g-r-3.6$\mu$m image of the cluster.  The contours result from the  filtering of Fig. \ref{cluster-image} by a wavelet adapted to low-count statistics \citep{starck98}: the image is de-noised but intensity is not strictly conserved. The first contour is indicative of the background level. The two green circles indicate the X-ray sources for which a spectroscopic redshift is available, in addition to source A (see Table \ref{pointsource-x}). The red circles highlight the B sources. The yellow-dashed circle has a radius of 1 Mpc at the cluster redshift. The yellow cross indicates the center of the IRAC overdensity (\cite{papovich10}).}
              \label{overlay}
    \end{figure*}

IRC-0218a was observed for 84.5 ks on 27-09-2010 with Chandra ACIS-S
(obsid 12882). The data were reduced following the standard procedures
using the Chandra Interactive Analysis of Observations CIAO version 4.3  and CALDB version 4.4.2 \citep{fruscione06}. After cleaning and filtering, the total useful time was 83.7 ks.  The photon image of the cluster field is  presented on Fig. \ref{cluster-image}. 
At the position of the main cluster and of the companion, bright point sources are detected while no extended emission is visible. 
For comparison, the XMM image is shown in  Fig. \ref{xmm-image}a. The XMM sources all show a tangential elongation typical of the XMM PSF at large off-axis angle, suggesting that they are indeed mainly point-like. 
The Chandra image overlaid on a g-r-3.6$\mu$m composite is displayed in Fig. \ref{overlay}.

Table \ref{pointsource-x} gives the list of detected point sources along with redshift.
Point source A  (row  = 12) is detected with 32 ($\pm 6 $) photons in a 10 arcsec aperture in the [0.5-2] keV band, which corresponds to a flux  of 1.6 ($\pm 0.4$) $10^{-15}$ \flux . The galaxy associated with point source A has spectroscopic redshift of 1.623 and is found to be the object closest to the center of the near-infrared galaxy overdensity. The rest-frame optical spectrum from \cite{tanaka10}  shows no indication of activity (no emission lines of any kind); its infrared colors  are consistent with a passive massive galaxy. Its hardness ratio is fairly soft (0.53).

Point sources B1 (row = 7) and B2 (row = 9) have a [0.5-2] keV flux of 8.5 ($\pm 5 $) $10^{-16}$ \flux\   and 5.1 ($\pm 0.6$) $10^{-15}$ \flux , respectively.
There are only photometric redshifts available for the galaxies associated with these sources: 1 ($\pm 0.1 $) and 0.5 ($\pm 0.1 $) for B1 and B2 respectively, which makes them unlikely cluster members.  The position of the ``companion cluster'' mentioned by \cite{tanaka10} on the basis of their analysis of the XMM data  appears to surround sources B1 and B2.  As is conspicuous in Fig. \ref{xmm-image}b, the XMM emission is located at extreme off-axis angle, where the PSF is highly distorted;  the XMM emission is probably entirely resolved into the two  Chandra sources with photometric redshift  at $z\lesssim 1$.   This suggests that the hypothesis proposed by \cite{tanaka10} that this is a second cluster at the same redshift as IRC 0218A is most likely invalid.

Object row = 10 (z=1.6487) has strong emission features, including MgII 2800, suggesting an AGN.  
Object row = 14 (z=1.6240) is the strongest IR 24 micron source in the cluster vicinity.  It appears to be a merger, and likely an AGN as well.

\begin{table*}[h]
\tiny
\begin{tabular}{cccccccccc}
\hline
ID/Classification  &  Ra    &  Dec    &  PSF Size (soft) &  PSF Size (hard)    &  Flux soft               &  CR soft            & HR &  Zspec      &  Zphot \\
-           &  deg &  deg    &  pix        &  pix           &  $10^{-15}$ ergs/cm$^2$/s & $10^{-4}$ cps            &  - &  -          &  -  \\
(1) &(2) &(3) & (4) & (5) & (6) & (7) & (8) & (9) &(10)  \\
\hline  
$      1$  &  $34.5809$  &  $-5.2543$  &  $3.51$  &  $-999.00$           &  $0.88_{-0.26}^{+0.28}$  &  $2.12_{-0.64}^{+0.68}$   &  $-0.217 \pm 0.308$ &  -  &  $1.38_{-0.43}^{+0.00}$ \\
$      2$  &  $34.5961$  &  $-5.2367$  &  $2.55$  &  $-999.00$           &  $0.94_{-0.29}^{+0.31}$  &  $2.26_{-0.70}^{+0.76}$   &  $-0.377 \pm 0.352$ &  -  &  $3.19_{-0.66}^{+0.01}$ \\
$      3$  &  $34.6173$  &  $-5.2326$  &  $2.47$  &  $2.45$           &  $0.47_{-0.24}^{+0.27}$  &  $1.13_{-0.59}^{+0.65}$   &  $0.368 \pm 0.289$ &  -  &  $2.37_{-0.10}^{+0.10}$ \\
$      4$  &  $34.6437$  &  $-5.2325$  &  $2.92$  &  $2.92$           &  $11.92_{-0.79}^{+0.80}$  &  $28.68_{-1.91}^{+1.93}$   &  $-0.389 \pm 0.057$ &  -  &  $0.69_{-0.12}^{+0.02}$ \\
$      5$  &  $34.5356$  &  $-5.2068$  &  $2.26$  &  $2.26$           &  $12.11_{-0.80}^{+0.81}$  &  $29.15_{-1.93}^{+1.94}$   &  $-0.512 \pm 0.057$ &  -  &  $2.23_{-1.54}^{+-0.69}$ \\
$      6$  &  $34.6091$  &  $-5.2013$  &  $1.23$  &  $-999.00$           &  $0.39_{-0.23}^{+0.26}$  &  $0.93_{-0.56}^{+0.62}$   &  $-1.000$ &  -  &  $1.47_{-0.08}^{+0.11}$ \\
$   7$ (B1)  &  $34.6386$  &  $-5.1977$  &  $1.54$  &  $-999.00$           &  $0.85_{-0.27}^{+0.29}$  &  $2.05_{-0.66}^{+0.71}$   &  $-0.641 \pm 0.359$ &  -  &  $1.01_{-0.14}^{+0.18}$ \\
$      8$  &  $34.5532$  &  $-5.1898$  &  $1.37$  &  $-999.00$           &  $0.76_{-0.29}^{+0.32}$  &  $1.82_{-0.70}^{+0.76}$   &  $0.034 \pm 0.299$ &  -  &  $1.11_{-0.04}^{+0.11}$ \\
$   9$ (B2)  &  $34.6345$  &  $-5.1892$  &  $1.24$  &  $1.23$           &  $5.11_{-0.54}^{+0.56}$  &  $12.30_{-1.31}^{+1.34}$   &  $-0.756 \pm 0.112$ &  -  &  $0.53_{-0.08}^{+0.05}$ \\
$ 10$ (AGN)  &  $34.5717$  &  $-5.1849$  &  $0.90$  &  $0.90$           &  $2.69_{-0.42}^{+0.44}$  &  $6.48_{-1.00}^{+1.05}$   &  $-0.064 \pm 0.127$ &  $1.6487$ (a)  &  $1.73_{-0.11}^{+0.07}$ \\
$     11$  &  $34.5514$  &  $-5.1785$  &  $1.17$  &  $-999.00$           &  $1.13_{-0.31}^{+0.33}$  &  $2.72_{-0.74}^{+0.79}$   &  $-0.699 \pm 0.296$ &  -  &  $0.27_{-0.08}^{+0.09}$ \\
$   12$ (A)  &  $34.5879$  &  $-5.1758$  &  $0.62$  &  $0.62$           &  $1.60_{-0.34}^{+0.36}$  &  $3.84_{-0.82}^{+0.86}$   &  $-0.530 \pm 0.252$ &  $1.6230$ (b)  &  $1.62_{-0.09}^{+0.08}$ \\
$     13$  &  $34.5423$  &  $-5.1757$  &  $1.36$  &  $-999.00$           &  $0.39_{-0.23}^{+0.25}$  &  $0.93_{-0.56}^{+0.61}$   &  $0.235 \pm 0.391$ &  -  &  $1.93_{-0.28}^{+0.02}$ \\
$14$ (IR-24$\mu$/AGN)  
             &  $34.5997$  &  $-5.1739$  &  $0.58$  &  $-999.00$           &  $1.33_{-0.33}^{+0.35}$  &  $3.21_{-0.80}^{+0.85}$   &  $-0.812 \pm 0.268$ &  $1.6240$ (b)  &  $1.76_{-0.15}^{+0.12}$ \\
$     15$  &  $34.5713$  &  $-5.1647$  &  $0.61$  &  $-999.00$           &  $0.58_{-0.26}^{+0.28}$  &  $1.39_{-0.62}^{+0.68}$   &  $-0.341 \pm 0.465$ &  -  &  $1.71_{-0.11}^{+0.07}$ \\
$     16$  &  $34.6130$  &  $-5.1637$  &  $0.53$  &  $-999.00$           &  $0.20_{-0.19}^{+0.20}$  &  $0.48_{-0.45}^{+0.49}$   &  $0.443 \pm 0.463$ &  -  &  $2.37_{-0.52}^{+0.21}$ \\
$     17$  &  $34.6478$  &  $-5.1515$  &  $0.97$  &  $-999.00$           &  $1.01_{-0.31}^{+0.34}$  &  $2.44_{-0.75}^{+0.81}$   &  $-0.209 \pm 0.296$ &  -  &  $2.02_{-0.25}^{+0.07}$ \\
$     18$  &  $34.5667$  &  $-5.1499$  &  $0.58$  &  $0.58$           &  $1.20_{-0.31}^{+0.33}$  &  $2.88_{-0.75}^{+0.80}$   &  $-0.026 \pm 0.222$ &  -  &  $1.82_{-0.12}^{+0.15}$ \\
$     19$  &  $34.5872$  &  $-5.1332$  &  $0.43$  &  $0.43$           &  $0.77_{-0.28}^{+0.30}$  &  $1.85_{-0.67}^{+0.72}$   &  $0.343 \pm 0.202$ &  -  &  $1.47_{-0.05}^{+0.09}$ \\
$     20$  &  $34.5530$  &  $-5.1311$  &  $0.83$  &  $-999.00$           &  $0.21_{-0.18}^{+0.20}$  &  $0.51_{-0.44}^{+0.48}$   &  $-0.012 \pm 0.704$ &  -  &  $0.58_{-0.02}^{+0.03}$ \\
$     21$  &  $34.6180$  &  $-5.2642$  &  $4.30$  &  $-999.00$           &  *                    &  *                     &  * &  -  &  $4.01_{-0.12}^{+0.10}$ \\
$     22$  &  $34.5492$  &  $-5.2329$  &  $2.96$  &  $-999.00$           &  $0.66_{-0.27}^{+0.29}$  &  $1.59_{-0.65}^{+0.70}$   &  $-0.849 \pm 0.530$ &  -  &  $0.01_{--0.01}^{+0.07}$ \\
$     23$  &  $34.5358$  &  $-5.2123$  &  $2.47$  &  $-999.00$           &  $0.57_{-0.24}^{+0.27}$  &  $1.37_{-0.58}^{+0.64}$   &  $-0.009 \pm 0.370$ &  -  &  $0.56_{-0.10}^{+0.05}$ \\
$     24$  &  $34.5255$  &  $-5.1546$  &  $1.64$  &  $-999.00$           &  $0.34_{-0.18}^{+0.20}$  &  $0.81_{-0.44}^{+0.49}$   &  $0.053 \pm 0.441$ &  -  &  $0.60_{-0.02}^{+0.03}$ \\
$     25$  &  $34.5583$  &  $-5.2059$  &  $-999.00$  &  $1.70$           &  $0.34_{-0.21}^{+0.23}$  &  $0.83_{-0.50}^{+0.55}$   &  $0.574 \pm 0.239$ &  -  &  $0.38_{-0.09}^{+0.02}$ \\
$     26$  &  $34.5965$  &  $-5.1205$  &  $-999.00$  &  $0.49$           &  $0.31_{-0.21}^{+0.23}$  &  $0.74_{-0.50}^{+0.55}$   &  $0.587 \pm 0.258$ &  -  &  $1.38_{-0.16}^{+0.05}$ \\
$     27$  &  $34.5269$  &  $-5.2332$  &  $-999.00$  &  $3.59$           &  $0.00^{+0.21}$          &  $0.00^{+0.51}$           &  $1.000$ &  -  &  $0.00_{-0.00}^{+0.00}$ \\
$     28$  &  $34.6478$  &  $-5.1635$  &  $-999.00$  &  $1.09$           &  $0.32_{-0.21}^{+0.23}$  &  $0.78_{-0.50}^{+0.55}$   &  $0.334 \pm 0.384$ &  -  &  $1.45_{-0.06}^{+0.07}$ \\
\hline
\hline
\end{tabular} 
\caption{ 
Catalog of point sources in the field. Columns (4) and (5) refer to the pixel size of the PSF at the location of the source as estimated by {\tt wavdetect}. A value of -999.9 indicates that the source was not detected in the given band. The soft and hard band correspond to [0.5-2] keV and [2-8] keV, respectively. Fluxes in column (6) are unabsorbed fluxes computed assuming a power law of index 1.7 and a galactic $N_{H}=2.2.10^{20} cm^{-2}$, using the count rates reported in column (7). Source 21 could not be measured because located at the edge of the FoV. If the mode of the flux (count rate) distribution is 0.0, the + sign indicates an upper limit. Fluxes are measured in 10 '' radius aperture. Column (8) provides the hardness ratio HR=(H-S)/(H+S) where H and S are count rates measured in hard and soft  bands, respectively.
Reference for the spectroscopic redhsifts: (a) \cite{papovich10}; (b) \cite{tanaka10}. The photometric redshifts are based on the same data and analysis as in \cite{papovich10}.
}
\label{pointsource-x} 
\end{table*}

\section{Spatial analysis}

\begin{figure}
   \centering      
        \includegraphics[width=8cm]{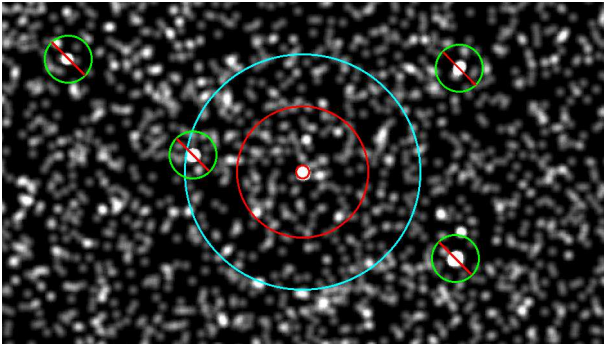}
   \caption{Chandra data ([0.3-2] keV band) around the cluster main component (A), smoothed by a Gaussian with a width of 2.5 arcsec.  The two red circles indicate the annulus ($2.5''- 25''$) in which  a possible 2.3 $\sigma$ extended contribution is detected; barred sources were masked in the analysis. Extending the search to $2.5''-45''$ (cyan circle) decreases the significance to below 1$\sigma$. }
              \label{profile-region}
    \end{figure}

\begin{figure}
   \centering      
        \includegraphics[width=8cm]{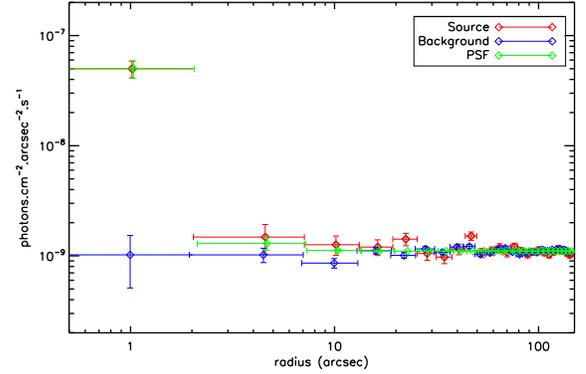}
   \caption{Surface brightness profile of the X-ray emission centered on source A.  The background is from the the Chandra blank-sky observations and was scaled to that of the present observation. The PSF model is from CIAO and was normalized to the central bin value. }
              \label{profile-A}
    \end{figure}

Despite the lack of conspicuous extended emission in the Chandra image, we statistically investigated the properties of the X-ray signal around source A.  For this purpose, we made use of command {\sc aprate\footnote{http://cxc.harvard.edu/ciao/ahelp/aprates.html}} in CIAO version 4.3. We use the  Chandra blank sky background files\footnote{ http://cxc.harvard.edu/ciao/threads/acisbackground/}. The exposure time of the blank-sky
    data was scaled so that the count rate in the image energy band
    agreed with the target data in source-free regions. This ensures
    that the differences in the soft Galactic foreground between the
    target and blank-sky fields do not impact our constraints on any
    possible extended emission. This exercise was performed for two energy bands, [0.5-2] keV and [0.3-2] keV within two annuli, $2.5''<r<25''$ and $2.5''<r<45''$, the lower bound allowing us to safely exclude source A\footnote{for the Chandra PSF, 95\% of the flux at 1.5 keV is within $2"`$   http://asc.harvard.edu/proposer/POG/html/chap4.html\#tth\_sEc4.2.3}; the other  point sources were removed as shown in Fig. \ref{profile-region}. A radius of $25''$ corresponds to 214 kpc at the cluster redshift, which is   about four times the scaled core radius (Sec. \ref{MASS}).  Within this annulus, we  detect a 2$\sigma$ and  2.3$\sigma$ signal in the [0.5-2] keV and [0.3-2] keV bands, corresponding to count rates, after vignetting correction, of $0.00025 \pm 0.00013$c/s and $0.00033 \pm 0.00015$c/s,   respectively. The [0.3-2] keV measurement shows a possible excess of  some 27 photons after a one-day observation. To detect a 3$\sigma$ signal in these conditions would have required 36 counts (scaled background value is 106 counts).
The mean signal detected in the annulus extending out to $45''$ is found to be below the one-sigma significance in both bands and is compatible with zero at the 3$\sigma$ level.    The [0.5-2 ] and  [0.3-2] count rates corresponds to  (absorbed) fluxes of    $1.1 ~10^{-15}$ and $3.3 ~10^{-15}$ \flux\ in $2.5''<r<25''$, respectively.  
    Results are summarized in Table \ref{xmes}  and the corresponding radial profile is shown in Fig. \ref{profile-A}

\begin{figure}
   \centering      
        \includegraphics[width=8cm, angle=0]{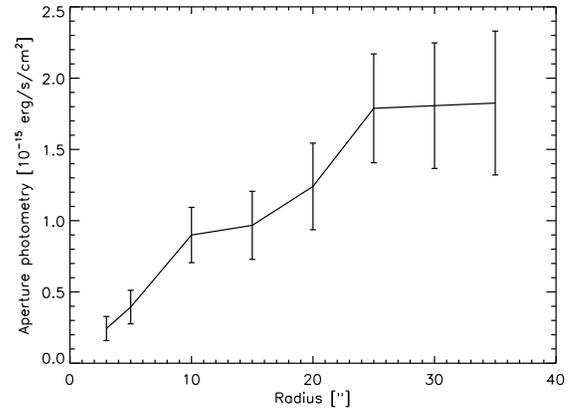}
   \caption{Growth curve of source A from the XMM image in the [0.5-2] keV band (dilution effect by the PSF is included, see text).}
              \label{xmm-grow}
    \end{figure}

We then compared  the Chandra data with the XMM observation, for which it is not possible to reliably exclude the point sources given the large off-axis angle. Fig. \ref{xmm-grow} suggests that all significant XMM emission is encompassed in a radius of $25''$. Accounting for the PSF dilution effect (0.9) in the $25''$ aperture, we measure an absorbed flux of $1.8^{+0.4}_{-0.4} ~10^{-15}$ \flux\ in the [0.5-2] keV band. For this calculation, we converted the measured XMM count rate  assuming an APEC plasma model defined by T= 3keV, Ab=0.3 and N$_{H}$=$2.2~10^{20}$ cm$^{-2}$. Integrating the Chandra data within the same radius (Fig. \ref{profile-region}) using {\sc aprate}  gives an absorbed flux of $2.6^{+0.6}_{-0.6}~10^{-15}$ \flux , which agrees with the XMM measurements, within the error bars; restricting the Chandra measurement to the point source itself  gives 1.6 ($\pm 0.4$) $10^{-15}$ \flux (Sec. 2);  we cannot exclude, however, the possibility that the central source is a variable AGN.  

In the following, we concentrate on the [0.3-2] band and on the $2.5''<r<25''$  annulus, which appear to yield the most significant detection.

A similar analysis could not be performed in the B region, because it  falls close to the Chandra detector  edge, but it also appears to be entirely dominated by the two point sources with photometric redshifts of $\sim$ 0.5 and 1.

\section{Hot gas content and mass estimates for IRC-0218}

\label{MASS}

The diffuse X-ray emission associated with IRC-0218, if any,  appears to be very weak in our deep Chandra observation: about one source photon per hour was collected. One can use this to estimate the mass associated to the $2.3\sigma$  detection that we infer in the  [0.3-2] keV band. For this, we assumed that the surface brightness follows a $\beta$-profile defined by $\beta=0.5$  (appropriate for the group-size objects) along with a self-similar scaling $R_{c} = 180/h  \sqrt{(T/7)} /E(z)$ and $R_{c}= 0.2\times R_{500} $ \citep{clerc11}. Considering the L-T relation of \cite{pratt09} for non cool-core clusters and the M-T relation of \cite{arnaud05}, we derived a virial mass of $M_{200}= 7.7(\pm 3.8)~10^{13}M_{\odot}$ along with $R_{500} = 31''$; our inferred $R_{200}$ is 494 kpc (for $z=1.62$).

\begin{figure}
   \centering      
        \includegraphics[width=8cm, angle=0]{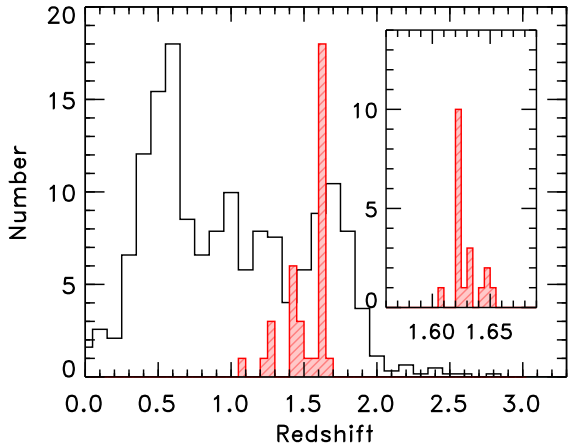}
   \caption{Redshifts available within 2 Mpc of the cluster core; the black  and red histograms  are for the photometric and spectroscopic redshifts, respectively. The inset panel shows a zoom of the redshift distribution around the cluster where the histogram uses bins of d(z)=0.005. }
              \label{zhist}
    \end{figure}

Fig. \ref{zhist} shows the distribution of all spectroscopically identified galaxies to date in the cluster region.  
There are currently 13 redshifts for galaxies with $1.62 < z < 1.65 $ within a physical projected radius of 1 Mpc off the cluster center \citep[10 of these galaxies have $1.62 < z < 1.63$;][; Momcheva et al. in prep; Willmer et al. in prep]{papovich10, tanaka10}.
Restricting  ourselves to those 10 galaxies and following the `gapper method' \citep{beers90}, we estimated a velocity dispersion of $360  \pm 90$ km/s where the error is derived using a jack-knife method; this value is substantially lower than the one quoted by \cite{papovich10}, because we exclude here galaxies above $z>1.63$.  From this, we estimated the  cluster virial mass following \cite{carlberg96}. The result is highly uncertain, formally $M_{200} = 2.2 (\pm 1.2)  10^{13} M_{\odot} $. 
The distribution of observed redshifts is subject to the following
selection biases: (1) successful spectroscopic redshifts are obtained
preferentially for the brighter galaxies in the cluster ; (2) because the
continuum is much fainter than the sky background, most of the redshifts
are obtained for galaxies with emission lines; (3) for objects at  $z >
1.625$, the identification of spectral features used for the redshift
determination is affected by the presence of strong sky emission lines.
Therefore,  the limited redshifts available for the cluster galaxies suggest that the virial mass is  lower than $3.4 ~ 10^{13} M_{\odot}$  (1 sigma), which is barely compatible with the lower limit inferred from the Chandra observation.

\begin{figure}
   \centering      
        \includegraphics[width=8cm, angle=0]{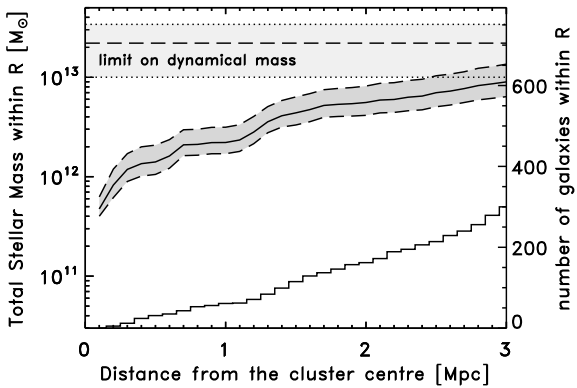}
   \caption{Stellar mass estimate as a function of distance from the center of the putative cluster. The solid curve shows the total stellar mass in all galaxies associated with the cluster (photometric redshifts); the gray region includes the 68\% confidence interval on the stellar mass for each galaxy.  The histogram shows the  total number of galaxies. The light-gray horizontal strip indicates our inferred 1$\sigma$ range for  the dynamical mass. }
              \label{stellarmass}
    \end{figure}

Finally, we provide an estimate of the stellar mass content of the putative cluster. This was computed from all  galaxies  around  $z \sim 1.6$,  i.e. galaxies having more than 40\% of their photometric redshift probability distribution function within $1.5 < z < 1.75$ (Fig. \ref{stellarmass}). 
We derived stellar masses by modeling the multiband Subaru (BViz), UKIDSS (JK), and IRAC (3.6-8.0 micron) photometry with a suite of stellar population models from Bruzual \& Charlot (2003).    We fitted the data for each galaxy with the stellar population models for a range of star-formation histories, stellar population ages, and dust extinction (all models assume solar metallicity and a Chabrier IMF; see \cite{papovich01} for details).
 The stellar mass  enclosed within the X-ray virial radius  ($\sim$ 500 kpc) accounts for  some 10 \%  of the reported estimated dynamical mass  and for a few percent  of  $M_{200}$ inferred from the X-ray data.

\begin{table*}
\caption{X-ray measurement summary for the central region. The last two lines include the central point source. Column {\em n(photons)} gives the number of source photons detected in the region defined by column 1, after subtraction of the scaled background; column {\em significance} gives the corresponding statistical significance of the detection.}            
\label{xdiffuse}       
\centering                          
\begin{tabular}{c c c cc c }       
\hline\hline                  Region & Band & n(photons) &significance &Flux & Inferred M200  \\   
 \hline                        
$ 2.5''-25''$  &0.5-2 keV & 21 &2.0 $\sigma$ &$1.1 \pm 0.5~10^{-15}$ (absorbed) &  \\       
 $2.5''-25''$ & 0.3-2 keV & 27 &2.3 $\sigma$ &$3.4 \pm 1.7~10^{-15}$ (absorbed) & 7.7$\times10^{13}M_{\odot} ~ \pm 50\%$   \\
$2.5''-45''$  &0.5-2 keV & 0.5 &-&$<1.2 ~10^{-15}$ (absorbed, 1$\sigma$ upper limit) &  \\       
 $2.5''-45''$ & 0.3-2 keV & 1 &- &$<3.3 ~10^{-15}$ (absorbed, 1$\sigma$ upper limit) &    \\
 \hline                                    
 $0-25''$  & 0.5-2 keV & &&$2.6 \pm0.6~10^{-15}$ (absorbed)  &   \\
 $ 0-25''$ & 0.5-2 keV & &&$1.8 \pm0.4~10^{-15}$ XMM (absorbed) &   \\
 \hline
\end{tabular}
\label{xmes}
\end{table*}

\section{Summary and conclusion}
The spatial analysis of our deep Chandra observation of IRC-0218A shows that the X-ray emission of this putative cluster of galaxies is entirely dominated by a point source coincident with a galaxy located at the centroid of the galaxy overdensity in the mid-infrared. The point source signal is consistent with the emission estimated from the XMM observation (at large off-axis).
The optical spectrum of this galaxy shows, however, no sign of activity and its X-ray hardness ratio is soft. 
We detected weak extended X-ray emission (2.3 $\sigma$) out to a radius of $25''$ (214 kpc) from the optical center. 
The inferred  virial mass corresponds to  a moderately massive cluster (5-10$\times 10^{13}M_{\odot}$) assuming that cluster scaling laws evolve self-similarly. Estimating a meaningful velocity dispersion for this object turned out to be very challenging: galaxies are faint and $z\sim1.6$ falls right in the ``redshift desert" where most
galaxy lines used to measure redshifts are displaced into the
near-infrared, where subtracting the OH emission and H2O atmospheric
absorption can potentially bias the ability to measure redshifts.
 There are moreover no clues about the degree of dynamical relaxation of the object.
Our current velocity estimate derived from 10 galaxies provides an upper limit for the dynamical mass that is marginally compatible with the  lower mass limit inferred from the Chandra data. The stellar mass estimate accounts for a few percent of the Chandra mass.
Although uncertainties are large, our Chandra observation along with the existing optical and IR data suggests that IRC-0218A is indeed a cluster or a collapsing cluster, rather than a filament seen in projection.

If the photo-z associated with the B1 and B2 sources are assumed to be correct, it follows that the X-ray twin-cluster hypothesis proposed by \cite{tanaka10} needs to be discarded.

More generally, we note that the X-ray emission of all four $z>1.5$ known clusters appears to be significantly  affected by point sources \citep[there is no Chandra image published yet for][ but the XMM images look very compact]{santos11, fassbender11}. Furthermore, clusters like IRC-0218A will certainly always escape direct X-ray detection in the XMM or Chandra archive for techniques solely based on the search for extended X-ray emission. This severely questions  the reliability of the determination of cluster scaling laws at high redshifts because the samples are likely to be severely biased toward objects that are over-luminous with respect to the mean.

\begin{acknowledgements}
    This research has made use of data obtained from the Chandra  Guest Observation n. 12882, and software provided by the Chandra X-ray Center (CXC) in the application packages CIAO. We thank Maxim Markevich for useful advices during the programing of the observation. CNAW acknowledges partial support from Chandra award G01-12157B .   \end{acknowledgements}

\end{document}